\documentclass[twocolumn,showpacs,preprintnumbers,amsmath,amssymb]{revtex4}
\usepackage{graphicx}
\usepackage{dcolumn}
\usepackage{bm}

\begin{document}
\title{Possible new electric effect in superfluid systems}
\author{S. I. Shevchenko}
\author{A. M. Konstantinov}
\email{shevchenko@ilt.kharkov.ua}
\affiliation{B. Verkin Institute for Low Temperature Physics and Engineering, Kharkov 61103, Ukraine}

\begin{abstract}
We predict that propagation of third sound in a superfluid film creates an electric field in the surrounding area that can be observable with modern measurement devices when there are many vortex pairs in the system (e. g. in the vicinity of the superfluid transition temperature).
\end{abstract}

\maketitle

Results of experiments on electric activity of He II were presented in \cite{1,2}. A standing third sound wave in He II was accompanied by appearance of electric potential about 10$^{-8}$\,V in the cell filled with helium \cite{1}. A potential difference of the same order appeared in a chamber covered with helium film during mechanical oscillations of the chamber \cite{2}. These experiments motivated a series of theoretical studies of polarization of normal and superfluid dielectric systems \cite{3}--\cite{12a}.

Discussing the possibility to observe theoretically predicted effects, we would like to note that there are two main problems. The first problem is related to the fact that small polarizability of helium atoms will make all of these effects small. The second problem is related to the nature of the polarization mechanism in the absence of an external electric field. When this electric field is absent, the source of polarization is the interaction between the atoms of the system. The ``building blocks'' of polarization are the dipole moments that pairs of atoms attain as a result of mutual van der Waals interaction. As is well known (see, for instance, \cite{13}), a pair of identical atoms get equal and oppositely directed dipole moments, the value of which depends on the distance between the atoms. The total dipole moment of the pair is equal to zero. In a non-uniform system we can get a redistribution of these dipole moments across the system, leading to local polarization. The total dipole moment of the system (in the stationary case) must be equal to zero. The local values of polarization can be observed by some kind of external ``probe''. An example of this ``probe'' might be a flux of electrons whose scattering will give us information about the system's state.

The present article deals with the possibility of a new electric effect in superfluid systems. The prediction is as follows: during the propagation of the third sound in a superfluid film, an electric field will appear in the surrounding space. This electric field will be observable if the system contains a large number of vortex pairs.

Let’s consider a thin superfluid film with a third sound propagating along it. As is known \cite{14,15,16}, the third sound is a surface wave of long wavelength on a liquid helium film during which the normal component remains stationary and the superfluid component oscillates parallel to the wall.  The liquid in the film can be considered incompressible. Therefore, in the third sound the film thickness $h$ will oscillate as a function of coordinates and time. Due to the motion of the superfluid component, the density of the normal component ${{\rho }_{n}}(\mathbf{r},T)=\rho -{{\rho }_{s}}(\mathbf{r},T)$ (${{\rho }_{n}}$, ${{\rho }_{s}}$ and $\rho $ are the two-dimensional normal, superfluid and total mass density, respectively), and with it the film temperature, must also oscillate in time and space. However, the changes in temperature are set not only by the motion of the superfluid component but also by heat exchange with the substrate. When films are thin enough, the heat will quickly escape into the substrate (see \cite{14,15,17} for details). As a result, the changes of the film's temperature can be neglected. We will also neglect film evaporation. Then, the state of the system can be describe by the superfluid component's equation of motion

\begin{equation} \label{1}
\frac{\partial {{\mathbf{v}}_{s}}}{\partial t}+({{\mathbf{v}}_{s}}\nabla ){{\mathbf{v}}_{s}}=-f\nabla h+\frac{{{\rho }_{n}}}{2\rho }\nabla {{\left( {{\mathbf{v}}_{n}}-{{\mathbf{v}}_{s}} \right)}^{2}},
\end{equation}
and the continuity equation, modified for the case at hand,
\begin{equation}\label{2}
\frac{\partial h}{\partial t}+\frac{{{\rho }_{s}}}{\rho }\mathrm{div}\left( h{{\mathbf{v}}_{s}} \right)=0.
\end{equation}
Here, $h(\mathbf{r},t)$ is the height of the film as a function of the two-dimensional coordinate $\mathbf{r}$ and time $t$, ${{\mathbf{v}}_{s}}(\mathbf{r},t)$ is the total superfluid velocity, defined both by potential fluxes ($\nabla \times {{\mathbf{v}}_{s}}=0$) and by vortices ($\nabla \cdot {{\mathbf{v}}_{s}}=0$), while ${{\mathbf{v}}_{n}}$ is the velocity of the normal component, which coincides with the velocity of the substrate, which we take to be immobile hereafter. The term proportional to $f$ is the van der Waals force of the substrate acting on the film.

In the absence of the third sound the height of the film ${{h}_{0}}$ does not depend on coordinates and time and the surface of the film (in the case of a planar substrate) is just a plane parallel to the substrate. It is important to note that perpendicular to the film dipole moment ${{\mathbf{p}}_{\bot }}$ appears near the surface of the film. As was mentioned above, due to interaction two identical atoms will attain dipole moments that have equal magnitude but that are oppositely directed along the line that connects their centers. If the distance between atoms is significantly greater than the Bohr radius $a_{B}$, the value of the dipole moment will be $A/{{R}^{7}}$, where $R$ is the distance between atoms and $A$ is a coefficient that depends on the atom type. If you place a chain of equidistant atoms between atoms of the same type, then (assuming that only the closest neighbors interact) the dipole moment of each internal atom will be equal to zero due to the mutual compensation of dipole moments of adjacent atoms. The dipole moments of the boundary atoms will be equal and oppositely directed. The appearance of a dipole moment on the surface of the film can be explained in a similar way.

Indeed, let's consider an atom which is located inside of a film at a distance $z$ from the substrate. We will assume that ${{h}_{0}}-z\ll {{h}_{0}}$. The dipole moment of this atom is obtained by adding up the moments induced by all of the other atoms. The component of the atom’s dipole moment that is perpendicular to the surface is equal to
\begin{equation}\label{3}
{{\mathbf{p}}_0} = {\mathbf{n}}\int {{d^2}r} \left( {\int\limits_a^z {dz' - \int\limits_a^{{h_0} - z} {dz'} } } \right)\frac{{z'}}{R}\frac{A}{{{R^7}}}{n_0}.
\end{equation}
Here, $R=\sqrt{{{r}^{2}}+{{{{z}'}}^{2}}}$, ${{n}_{0}}$ is the equilibrium three-dimensional atom density, $a$ is a length scale that is of order the distance between atoms and $\mathbf{n}$ is a unit vector directed along the normal to the film surface. Formula \eqref{3} takes into account the fact that the perpendicular component of the dipole moment of the atom in the point $(0,0)$, which is induced by an atom in the point $(r,{z}')$, is obtained by multiplying the dipole moment $A/{{R}^{7}}$ by the cosine of the angle between the vector $\mathbf{n}$ and the vector that connects these atoms. The right hand side of \eqref{3} is written as a difference of integrals. The first integral takes into account the interaction of the analyzed atom with atoms that lie below it while the second integral takes into account the interaction with the atoms above it. Carrying out the integration and multiplying the result by the atom density ${{n}_{0}}$, we get the polarization vector of the plane near its surface
\begin{equation}\label{4}
{{\mathbf{P}}_{0}}(z)=\frac{\pi n_{0}^{2}A}{12{{({{h}_{0}}-z)}^{4}}}\mathbf{n}.
\end{equation}
This expression shows that the polarization drops rapidly as we get farther away from the surface. For this reason, we can assume that the film gets a surface polarization ${{\mathbf{P}}_{S}}$, the expression for which can be obtained by integrating \eqref{4} with respect to $z$ from the film’s bottom up to its surface. As a result, we find that
\begin{equation}\label{5}
{{\mathbf{P}}_{S}}=\frac{\pi n_{02}^{2}A}{36{{a}^{5}}}\mathbf{n}=e{{a}_{B}}\frac{\pi }{2}{{\left( \frac{{{a}_{B}}}{a} \right)}^{7}}n_{02}\mathbf{n}\equiv {{P}_{S}}\mathbf{n},
\end{equation}
where we have introduced the equilibrium two-dimensional density of atoms ${{n}_{02}}$ with the help of the relation ${{n}_{02}}={{n}_{0}}a$ and took into account the fact that for helium atoms, $A\approx 18.4ea_{B}^{8}$.

The surface dipole moment ${{\mathbf{P}}_{S}}$ will lead to a non-zero electric potential above the film, which will be almost unobservable due to its small value and constant-in-time nature. The situation is changed if a third sound propagates in the film. During oscillations of the surface the normal to the surface at a given point will attain a component parallel to the substrate which, in turn, will lead to the appearance of an atomic dipole moment component that is parallel to the substrate. When the inclination angles of the normal $\mathbf{n}$ are small, this component is equal to $-{{p}_{0}}\nabla h(\mathbf{r},t)$. This dipole moment will, in the presence of third sound propagation, lead to the appearance of an electric field $\mathbf{E}(\mathbf{r},t)$ in the surrounding space. The field $\mathbf{E}(\mathbf{r},t)$ is proportional to $\nabla h$, while the derivative $\nabla h\sim{{h}_{0}}\mathbf{k}$, where $\mathbf{k}$ is the wave vector of the surface wave. We will be interested in the case where the wavelength that corresponds to the wave vector is $\lambda \sim 1$\,cm. Then, ${{h}_{0}}/\lambda \sim{{10}^{-7}}...{{10}^{-6}}$.  Since $\mathbf{E}\sim\nabla h$, in the absence of vortices, this field is so small that observations are impossible. The presence of vortices whose densities near the transition of the film into a superfluid state can reach ${{10}^{11}}-{{10}^{13}}$\,cm$^{-2}$ will increase the strength of the electric field by many orders of magnitude. Let's calculate the value of the field that appears above a film in the presence of vortices.

We are interested in the vortices' contribution to the deformation of the film surface, which will occurs during the propagation of the third sound. A wave, incident on a vortex, will scatter off of it, and the scattered wave will lead to deformations of the surface. The full velocity is ${{\mathbf{v}}_{s}}={{\mathbf{v}}_{v}}+\mathbf{{v}'}$ where ${{\mathbf{v}}_{v}}$ is the velocity related to the vortex and $\mathbf{{v}'}={{\mathbf{v}}_{1}}+{{\mathbf{v}}_{2}}$ where ${{\mathbf{v}}_{1}}$ is the velocity of the incident wave and ${{\mathbf{v}}_{2}}$ is the velocity of the scattered third sound wave (compare with \cite{18}). If the vortex moves as a whole, then in the case of a planar substrate, ${{\bf{v}}_{{v}}} = (\hbar /m)({\bf{n}} \times ({\bf{r}} - {{\bf{v}}_L}t)/\left| {{\bf{r}} - {{\bf{v}}_L}t} \right|)$. The velocity of the vortex is ${{\mathbf{v}}_{L}\equiv\dot{\bf{r}}}$. We will assume that the velocity of the fluid in the incident wave is equal to ${{\mathbf{v}}_{1}}(\mathbf{r},t)=(\mathbf{k}/k){{v}_{1}}\exp \left[ i\left( \mathbf{k}\cdot \mathbf{r}-\omega t \right) \right]$ and ${{h}_{1}}(\mathbf{r},t)$ and ${{h}_{2}}(\mathbf{r},t)$ are the changes of the film's height caused by the incident and scattered wave respectively.

Assuming that the scattered wave is weaker than the incident wave, we obtain from \eqref{1} and \eqref{2} that
\begin{multline}\label{7}
\frac{\partial {{\mathbf{v}}_{2}}}{\partial t}+f\nabla {{h}_{2}}=-\frac{\partial {{\mathbf{v}}_{v}}}{\partial t}-\nabla \left( {{\mathbf{v}}_{1}}\cdot {{\mathbf{v}}_{v}} \right)+\frac{{{\rho }_{n}}}{\rho }\nabla \left( {{{\mathbf{v}}_{1}} \cdot {{\mathbf{v}}_{v}} } \right),
\end{multline}
\begin{equation}\label{8}
\frac{\partial {{h}_{2}}}{\partial t}+\frac{{{\rho }_{S}}}{\rho }{{h}_{0}}\mathrm{div}{{\mathbf{v}}_{2}}=-\frac{{{\rho }_{s}}}{\rho }({{\mathbf{v}}_{v}}\cdot \nabla ){{h}_{1}}.
\end{equation}

From this system it is easy to find a closed equation for ${{h}_{2}}(\mathbf{r},t)$. But first we note that for a vortex moving as a whole $\partial \mathbf{v}_v/\partial t=\nabla(\hbar/m)\partial\phi(\mathbf{r}-\mathbf{r}_v(t))/\partial t = -\nabla(\hbar/m)(\partial\phi/\partial\mathbf{r})\cdot \dot{\mathbf{r}}_v=-\nabla(\mathbf{v}_v\cdot\mathbf{v}_L)$. Also taking into account the fact that the scattered waves, like the incident waves, are proportional to $\exp\left[i(\mathbf{k}\cdot\mathbf{r}-\omega t)\right]$ in such a way that $\partial {{h}_{2}}/\partial t\sim{{c}_{3}}k$ and $\partial{{\mathbf{v}}_{2}}/\partial t\sim{{c}_{3}}\mathbf{k}$, where $c_3=(\rho_s h_0 f/\rho)^{1/2}$ is the third sound velocity \cite{14}, we get from \eqref{7} and \eqref{8} the equation for the deformation of the film surface in the presence of a vortex
\begin{multline}\label{9}
{{\nabla }^{2}}{{h}_{2}}+{{k}^{2}}{{h}_{2}}={{f}^{-1}}\left[ \frac{{{\rho }_{s}}{{k}^{2}}}{\rho }\left( {{\mathbf{v}}_{1}}\cdot {{\mathbf{v}}_{v}} \right)+\frac{{{\rho }_{n}}}{\rho }{\nabla^2} \left( {{{\mathbf{v}}_{1}} \cdot {{\mathbf{v}}_{v}} } \right)-\right.\\\left.-i\frac{{{\rho }_{s}}k}{\rho {{c}_{3}}}\left( {{\mathbf{v}}_{1}}\cdot \nabla  \right)\left( {{\mathbf{v}}_{L}}\cdot {{\mathbf{v}}_{v}} \right)+{{\nabla }^{2}}\left( \left( {{\mathbf{v}}_{L}}-{{\mathbf{v}}_{1}} \right)\cdot {{\mathbf{v}}_{v}} \right) \right].
\end{multline}
We can at ones write down the solution of this equation, but it makes more sense to simplify it first. As will be shown later, the velocity of the vortex   is determined by the velocity of the incident wave: ${{{v}}_{L}}\sim{{{v}}_{1}}$. Therefore, the term on the right hand side of \eqref{9} that is proportional to $k$ is of order ${v}_{1}^{2}$ and, since we have a linear approximation in $v_1$, it is pointless to keep this term around. All terms proportional to $k^2$ can also be neglected, since (as we will soon show), the electric field that appears above the film is proportional to $k$. The solution of the simplified equation has the form
\begin{equation}\label{10}
{{h}_{2}}(\mathbf{r})={{f}^{-1}}\left[ \left( {{\mathbf{v}}_{L}}-{{\mathbf{v}}_{1}} \right)\cdot {{\mathbf{v}}_{v}}+\frac{{{\rho }_{n}}}{\rho }{{\mathbf{v}}_{1}} \cdot {{\mathbf{v}}_{v}} \right].
\end{equation}
The velocity ${{v}_{L}}$ can be found from the equation of balance of the forces that act on the vortex (see, for instance, \cite{17,19,20})
\begin{multline}\label{11}
{{\rho }_{s}}\kappa \left[ ({\mathbf{v}_{L}}-{\mathbf{v}_{1}})\times \mathbf{n} \right]=-{{D}_{1}}{\mathbf{v}_{L}}-\frac{\kappa }{\left| \kappa  \right|}{{D}_{2}}\left[ \mathbf{n}\times {\mathbf{v}_{L}} \right],
\end{multline}
where $\kappa =2\pi \hbar /m$ is the circulation of the vortex. The expression on the left hand side is the Magnus force. The first term on the right hand side is friction while the second term is the Iordanskii force. The coefficients ${{D}_{1}}$ and ${{D}_{2}}$ are determined by ripplons scattering off of the vortex. Solving the force balance equation, it's easy to find the velocity of the vortex.

When the temperature of the film  is close to the temperature of transition into the superfluid state ${{T}_{BKT}}$ (${{T}_{BKT}}$ is the Berezinskii–Kosterlitz–Thouless transition temperature), the film will contain a large number of vortex pairs with opposite circulations \cite{21,22}. Therefore, it makes sense for us to now focus on finding the surface deformation caused by a vortex pair. The velocity of the motion of each component of the pair is still defined by equation \eqref{11}, where we must add to ${{\mathbf{v}}_{1}}$ (which is the velocity of the external flux) the velocity of the flux created near the given vortex by the other component of the pair. The velocity of this flux is equal to ${{\mathbf{v}}_{l}}=({\hbar }/{m{{l}^{2}})}\left[ \mathbf{n}\times \mathbf{l} \right]$, where $\mathbf{l}$ is the radius-vector from the vortex with positive circulation to the vortex with negative circulation. With the help of \eqref{10} and \eqref{11}, we find that
\begin{equation}\label{13}
{{h}_{2}}(\mathbf{r})=-\frac{2\left| \kappa  \right|{{\rho }_{s}}D}{Tf}\left( \mathbf{u}(\mathbf{l})\cdot {{\mathbf{v}}_{v}}(\mathbf{r}) \right).
\end{equation}
Here, ${{\mathbf{v}}_{v}}$ is the velocity created by the vortex with positive circulation. We have also introduced a coordinate-independent velocity $\mathbf{u}(\mathbf{l})=\mathbf{n}\times ({{\mathbf{v}}_{1}}+{{\mathbf{v}}_{l}})$. The diffusion coefficient $D$ in \eqref{13} can be easily written out in terms of the coefficients ${{D}_{1}}$ and ${{D}_{2}}$. According to estimates in \cite{17}, near $T_{BKT}$ this coefficient is equal to $D=\hbar /m$.

Assuming that the height ${{h}_{2}}(\mathbf{r})$ is known, we can find the electric potential above the film. In a point with coordinates ${{\mathbf{R}}_{0}}=({{\mathbf{r}}_{0}},{{z}_{0}})$, the potential created by the vortex pair is equal to
\begin{equation}\label{14}
\phi ({{\mathbf{R}}_{0}})=-{{P}_{S}}\int{\nabla {{h}_{2}}\cdot \nabla \frac{1}{\left| \mathbf{R}-{{\mathbf{R}}_{0}} \right|}}{{d}^{2}}r.
\end{equation}
We take into account the fact that the dipole moment ${{P}_{S}}\nabla {{h}_{2}}$ lies in the plane of the film (it has only $x$ and $y$ components) and integrate over the surface of the film. We use the symbols $\left| \mathbf{R}-{{\mathbf{R}}_{0}} \right|={{\left[ {{\left( \mathbf{r}-{{\mathbf{r}}_{0}} \right)}^{2}}+{{(h(\mathbf{r})-{{z}_{0}})}^{2}} \right]}^{1/2}}$ and $\nabla =\partial /\partial \mathbf{r}$. We are interested in electric fields at distances from the film that are large compared to its thickness (as in ${{z}_{0}}\gg h(\mathbf{r})$). This will allow us to neglect $h(\mathbf{r})$ in $\left| \mathbf{R}-{{\mathbf{R}}_{0}} \right|$. After substituting to \eqref{14} the expression for ${{h}_{2}}(\mathbf{r})$ from \eqref{13} we obtain
\begin{equation}\label{15}
\phi ({{\mathbf{R}}_{0}})=C\left[ \mathbf{u}(\mathbf{l})\times \mathbf{n} \right]\cdot \frac{({{\mathbf{r}}_{0}}-{{\mathbf{r}}_{{v}}})}{{{\left[ {{({{\mathbf{r}}_{0}}-{{\mathbf{r}}_{{v}}})}^{2}}+z_{0}^{2} \right]}^{3/2}}},
\end{equation}
where $C={{P}_{S}}{{\kappa }^{3}}{{\rho }_{S}}/2\pi Tf$, and ${{\mathbf{r}}_{{v}}}$ is the symbol for the coordinate of the vortex pair. The position of the pair can be described by the coordinate ${{\mathbf{r}}_{{v}}}$ when the distances from the components of the pair to the point of observation are large compared to the size of the pair itself. This is the case that we are interested in. From \eqref{15}, it follows that the electric potential in the point  coincides with the potential created by the dipole moment
\begin{equation}\label{16}
{{\mathbf{p}}_{pair}}(\mathbf{l})=C\left[ \mathbf{u}(\mathbf{l})\times \mathbf{n} \right].
\end{equation}
Since $\mathbf{u}(\mathbf{l})$ is a function of $\mathbf{l}$,this dipole moment depends on the size of the pair and its spatial orientation. We are interested in the potential $\phi ({{\mathbf{R}}_{0}})$, which is created by all of the vortex pairs that appear in the film due to thermal fluctuations. We will assume that the system is outside of the temperature range for which ${{T}_{BKT}}-T\ll {{T}_{BKT}}$. Then, the vortex pairs can be considered to be non-interacting with each other, and the probability of finding a pair on the area $R^2$ is proportional to the sum of the exponents $\exp \left[ -{{E}_{v}}\left( \left| {{\mathbf{r}}_{+}}-{{\mathbf{r}}_{-}} \right|/\xi  \right)/T \right]$ with respect to all of the ${{\mathbf{r}}_{+}}$ and ${{\mathbf{r}}_{-}}$ (which are the coordinates of vortices with positive and negative circulation, respectively) that lie on this area. The energy of a pair, in the presence of a relative motion between the normal and superfluid components with a velocity  $\mathbf{w}={{\mathbf{v}}_{n}}-{{\mathbf{v}}_{s}}$, is equal to (see, for instance, \cite{23})
\begin{multline}\label{17}
{E_v} = \frac{{ 2\pi \rho_s \hbar}}m {\left( {\frac{\hbar }{m}} \ln \frac{l}{\xi }  + {\mathbf{l}} \cdot \left[ {{\mathbf{n}} \times {\mathbf{w}}} \right]\right)}+ 2\Delta,
\end{multline}
where $\Delta$ is the energy of the core of the vortex and ${\mathbf{l}} = {{\mathbf{r}}_- } - {{\mathbf{r}}_+ }$. The cutoff length $\xi $ allows us to avoid divergences of energy on small scales. Typically this value is understood to be the radius of the vortex core. In the Kosterlitz–Thouless theory (which we follow), $\xi$ is also the step of the two-dimensional lattice, the knots of which may house the vortex cores. The summation with respect to ${{\mathbf{r}}_{+}}$ and ${{\mathbf{r}}_{-}}$ can be transformed into a double integral and all of the possible pair configurations can be counted. An area of ${{R}^{2}}$ has ${{(R/\xi )}^{2}}$ spaces for one component of the pair. The number of possible positions of the second component inside the ``volume'' element   is equal to ${{d}^{2}}l$ is $({{d}^{2}}l/{{\xi }^{2}})$. Then, the probability of finding a pair on the area ${{R}^{2}}$ and in the volume element ${{d}^{2}}l$ is equal to $dP({\mathbf{l}}) = {(R^2/{\xi ^2})}\exp \left[ { - {E_v}({\mathbf{l}})/T} \right]{d^2}l/\xi^2$. The electric potential in the point ${{\mathbf{R}}_{0}}$, which is created by all of the vortex pairs, is obtained from \eqref{15} by replacing the dipole moment of the pair \eqref{16} with the average moment
\begin{equation}\label{19}
\overline{\mathbf{P}}\equiv \int\limits_{\xi }^{\infty }{{{\mathbf{p}}_{pair}}}(\mathbf{l})\frac{d\mathbf{P}(\mathbf{l})}{{{R}^{2}}}=G(T){{n}_{v}}\mathbf{w},
\end{equation}
and integrating the result with respect to ${{d}^{2}}{{r}_{v}}$. In \eqref{19} ${{n}_{v}}$ is the number of pairs per unit area
\begin{multline}\label{20}
{n_v} \equiv \int\limits_\xi ^\infty  {\frac{{dP({\mathbf{l}})}}{{{R^2}}}} = \frac{\pi }{{{\xi ^2}}}\;\frac{T}{{(\pi \,{\hbar ^2}{\rho_s}/{m^2}) - T}}\exp [ - 2\Delta /T].
\end{multline}
Also in \eqref{19}, we have introduced the symbol $G(T)=C\left( {{\rho }_{s}}{{\kappa }^{2}}/4\pi T-1 \right)$. It is easy to see that $G({{T}_{BKT}})=4{{P}_{S}}\kappa /f$. While calculating the average of \eqref{19}, we assumed that in the energy expression \eqref{17}, the term proportional to the velocity $\mathbf{w}$ was small compared to the temperature.

Up until this point we have assumed that the substrate that the superfluid film covers is planar. However, it is important to take into account the fact that a superfluid film always spreads across the surface of a solid body, completely covering it (if the temperature of the body is less than ${{T}_{BKT}}$). As a result, the form of the body will have an influence on the result of the measurements. We will take, for our substrate, a cylindrical vessel with a radius ${{r}_{c}}$ and a height ${{L}_{z}}\gg {{r}_{c}}$. Therefore, we come to the problem of the propagation of the third sound in a film that covers the surface of a cylinder. We choose a cylindrical coordinate system and take the $z$ axis to be directed along the axis of the cylinder. We will also assume that the thickness of the film ${{h}_{0}}$ and the radius of the cylinder ${{r}_{c}}$ satisfy ${{r}_{c}}\gg {{h}_{0}}$. The peculiar aspect of the system at hand is its multiply connected nature. A system like this has no real BKT transition \cite{24}. However at large (compared to the average distance between vortices) cylinder radii, the system’s temperature of transition into a superfluid state practically coincides with ${{T}_{BKT}}$, and the behavior of the system of vortices only slightly differs from their behavior in planar systems. If, in addition to this, the radius of the cylinder is large compared to the distance from the point of observation to the cylinder (as in, ${{r}_{c}}\gg {{r}_{0}}-{{r}_{c}}$), then the formulas \eqref{15}...\eqref{20} stay valid.

Returning to the calculation of the electric potential, obtain with the help of \eqref{15}...\eqref{20}
\begin{equation}\label{21}
\phi ({{\mathbf{R}}_{0}}{)}=G{{n}_{v}}\int{\mathbf{w}(\mathbf{r})\cdot \nabla \frac{1}{\left| \mathbf{R}-{{\mathbf{R}}_{0}} \right|}{{d}^{2}}r}.
\end{equation}
We would like to point out that in the new coordinate system, $\left| \mathbf{R}-{{\mathbf{R}}_{0}} \right|\hm={{\left[ {{\left( {{\mathbf{r}}_{c}}-{{\mathbf{r}}_{0}} \right)}^{2}}+{{(z-{{z}_{0}})}^{2}} \right]}^{1/2}}$.We will assume that a standing wave regime is realized in the experiment. In this case, ${{h}_{1}}=\delta {{h}_{0}}\cos \omega t\sin kz$. This expression takes into account the fact that the third sound is excited by oscillations of the height of the film on the edges of the cylinder. In the future, we will assume that ${{L}_{z}}=\lambda /2\equiv \pi /k$. The velocity of the relative motion is equal to ${{{w}}_{z}}={{{w}}_{0}}(t)\cos kz$, where ${{{w}}_{0}}$ is given by the expression ${{{w}}_{0}}(t)=(\rho \delta {{h}_{0}}{{c}_{3}}/{{\rho }_{S}}{{h}_{0}})\sin \omega t$ \cite{14}. Putting this expression into \eqref{21}, we find that
\begin{multline}\label{22}
\phi {(}{{\mathbf{R}}_{0}}{)}=\\=G{{n}_{v}}{{{w}}_{0}}(t)\int{\cos kz\frac{\partial }{\partial z}\frac{1}{\left| \mathbf{R}-{{\mathbf{R}}_{0}} \right|}{{d}^{2}}r}=-G{{n}_{v}}{{{w}}_{0}}(t)\times\\\times\frac{\partial }{\partial {{z}_{0}}}\int{\frac{\cos kz}{\sqrt{r_{c}^{2}+r_{0}^{2}-2{{r}_{0}}{{r}_{c}}\cos \theta +{{(z-{{z}_{0}})}^{2}}}}}{{r}_{c}}d\theta dz.
\end{multline}
The potential $\phi ({{\mathbf{R}}_{0}})$ can be calculated analytically if $k{{r}_{c}}\ll 1$ and $k{{z}_{0}}\ll 1$. Without getting into details, for ${{r}_{0}}>{{r}_{c}}$ (e.g. outside the cylinder) we find that
\begin{equation}\label{23}
\phi ({{\mathbf{R}}_{0}})=-4\pi G{{n}_{v}}{{{w}}_{0}}(t)(k{{r}_{c}})(k{{z}_{0}})\ln \left( k{{r}_{0}} \right).
\end{equation}
This expression allows us to find the electric field as a function of coordinates. Differentiating \eqref{23} with respect to ${{r}_{0}}$ and ${{z}_{0}}$, we find that
\begin{equation}\label{24}
{{E}_{r}}{(}{{\mathbf{R}}_{0}}{)}=4\pi kG{{n}_{v}}{{{w}}_{0}}(t)k{{z}_{0}}\frac{{{r}_{c}}}{{{r}_{0}}},
\end{equation}				
\begin{equation}\label{25}
{{E}_{z}}{(}{{\mathbf{R}}_{0}}{) = }4\pi kG{{n}_{v}}{{{w}}_{0}}(t)k{{r}_{c}}\ln \left( k{{r}_{0}} \right).
\end{equation}

Constructed theory is valid also in the case when the superfluid film covers the inner surface of the hollow cylinder. The potential and the electric fields inside the cylinder can be obtained from \eqref{22} supposing that ${{r}_{0}}<{{r}_{c}}$. In this case
\begin{equation}\label{26}
\phi ({{\mathbf{R}}_{0}})=-4\pi G{{n}_{v}}{{{w}}_{0}}(t)(k{{r}_{c}})(k{{z}_{0}})\ln \left( k{{r}_{c}} \right),
\end{equation}
\begin{equation}\label{27}
{{E}_{r}=0,  {E}_{z}}{(}{{\mathbf{R}}_{0}}{) = }4\pi kG{{n}_{v}}{{{w}}_{0}}(t)k{{r}_{c}}\ln \left( k{{r}_{c}} \right).
\end{equation}

As a conclusion, let's get an estimate of the coefficient $G$ and the longitudinal electric field ${{E}_{z}}$ near the transition temperature for the case when ${{r}_{0}}>{{r}_{c}}$. Taking the formula $G=4{{P}_{S}}\kappa /f$ and plugging in the numerical values ${{P}_{S}}\approx e\cdot (30$\,cm$^{-1})$ and $\kappa =2\pi \hbar /m$, we find that $G\approx (e/f)(8\cdot {{10}^{-2}})$\,m/s. When $k\approx \pi$\,cm${^{-1}}$ and ${{n}_{v}}\approx {{10}^{12}}$\,cm${^{-2}}$, the amplitude of the longitudinal field is ${{E}_{z}}\approx {{f}^{-1}}({{10}^{5}}$\,V/s$^2)$. For unsaturated films, $f=\alpha /h_{0}^{4}$, where $\alpha $ is a parameter that characterizes the material composition of the substrate. For instance, for (CaF$_2$) (see, for example, \cite{25}), $\alpha \approx 2.2\cdot {{10}^{-14}}\,$erg$\cdot$cm$^{3}$/g. As a result, when ${{h}_{0}}=3\cdot {{10}^{-7}}$\,cm, we find that ${{E}_{z}}\approx 4\cdot {{10}^{-8}}$\,V/cm. The estimate for the field ${E}_{z}$ inside the cylinder remains the same as the field ${E}_{z}$ outside the cylinder because the expression \eqref{27} differs from \eqref{25} only by replacement of $\ln \left( k{{r}_{0}} \right)$ with $\ln \left( k{{r}_{c}} \right)$, but these logarithms are of the same order.

In the preceding analysis we have implicitly assumed that in the film of finite thickness, the vortices maintain a straight structure, with axes normal to the substrate. However, it appears plausible as the thickness of the film increases, the vortices and antivortices, which form correlated pairs, will start to ``close in'' on each other near the surface of the film, transforming into vortex semi-rings. Subsequent increases of film thickness will lead to these semi-rings transforming into vortex rings, whose axes are parallel to the substrate. We assume that for as long as the radius of typical vortex rings remain lower than the thickness of the film, the obtained results will be correct within an order of magnitude. All signs point to this thickness being of order a few hundred nanometers.

The ideas stated suggest that, during second sound propagation in the fluid, the vortex rings will get polarized, which will lead to electric fields appearing outside of the fluid. But that's a problem for another time.

\end{document}